\documentclass[a4paper,11pt]{article}
\pdfoutput=1
\usepackage{jinstpub}

\usepackage{graphicx}
\usepackage{amsmath}
\usepackage{hyperref }
\usepackage{lineno}
\usepackage{siunitx}

\graphicspath{ {figures/} }

\renewcommand{\vec}[1]{\mathbf{#1}}

\newcommand{\mrat}{\mathcal{R}}
\newcommand{\rat}{$\mathcal{R}$}

\title{Waveform resampling with LMN method}
\date{\today}

\author[a,b]{L. Gerlach}
\author[a]{W. Gu}
\author[a]{N. Nayak}
\author[a]{X. Qian}
\author[a,1]{B. Viren\note{Corresponding author.}}
\affiliation[a]{Brookhaven National Laboratory, Upton, NY 11973, USA}
\affiliation[b]{Princeton University, Princeton, NJ 08544, USA}
\emailAdd{bv@bnl.gov}

\abstract{
  Resampling is a common technique applied in digital signal processing.
  Based on the Fast Fourier Transformation (FFT), we apply an optimization
  called here the LMN method to achieve fast and robust re-sampling.  In
  addition to performance comparisons with some other popular methods, we
  illustrate the effectiveness of this LMN method in a particle physics
  experiment: re-sampling of waveforms from Liquid Argon Time Projection
  Chambers.
}


\begin{document}
\maketitle
\flushbottom

\section{Introduction}

Resampling is the process of transforming a signal in the form of samples collected at discrete points in some domain to a new set of samples collected at novel points in that domain.
The quality of a resampling is typically judged 
by how accurately these novel samples represent the original, underlying signal.
One central category of resampling methods is interpolation.
It asserts some model of the underlying signal and is normalized to preserve the local signal amplitude.
Other normalization policies may be asserted.
For example, an application may require the resampling to leave the integral of the signal invariant.

The computational expense of a particular software implementation of a resampling method is another important consideration.
Various software packages that are used in the analyses of data from particle physics experiments provide resampling method implementations including Boost~\cite{BoostLibrary}, Eigen~\cite{eigenweb}, SciPy~\cite{2020SciPy-NMeth} and ROOT~\cite{BRUN199781}.

These methods can be placed in two broad categories.
The first category operates in the sampled domain (e.g. interval or time domain)
and include
Cardinal B-spline interpolation~\cite{schoenberg1973cardinal}, Barycentric Rational interpolation~\cite{kong2021fastlinearbarycentricrational}, Hermite interpolation~\cite{cardinalhermiteinterpolation}, and Catmull Rom spline interpolation~\cite{catmull1974class}.
Interval domain methods can be fast but tend to not preserve spectral features.

The second category includes methods that consider the signal spectrum or otherwise operate in the frequency domain.
These include Whittaker-Shannon interpolation~\cite{whittaker1915interpolatory} and trigonometric interpolation~\cite{atkinson2009triginterp,kress1998triginterp}.
Typically, these methods apply operations in the frequency domain that are efficient while the discrete Fourier transform (DFT) as implemented with the Fast Fourier Transform (FFT) algorithm~\cite{cooley1965fft}\footnote{Gauss developed an algorithm akin to FFT in 1805 for a similar purpose of interpolating orbital measurements~\cite{heiderman1985gaussfft}.}
provide an efficient means to transform a signal between interval and frequency domain representations.
While the FFT is indeed fast, its complexity still grows as $\mathcal{O}(N\log N)$ in the size $N$ of the input signal.
Naive resampling methods can require inflating the original signal size by a large factor.
This inflation subjects the frequency domain method to ruinously large $N$.

Scaling problems can be exacerbated when the signal is large and multi-dimensional.
Interval-domain resampling can be extended to this case.
For example the bilinear interpolation~\cite{Kirkland2010} is a commonly employed method to interpolate images in the two-dimensional interval or pixel domain.
As with the one-dimensional case, these extensions to two-dimensional intervals tend to distort spectral features which can be problematic in some applications.
One dimensional methods that resample in the frequency domain may also be extended to multiple dimensions in a simple manner when those dimensions are independent.  They require application of a series of DFT along each dimension to reach their frequency domain.  An FFT applied to a two-dimensional signal of size $N \times M$ then costs $\mathcal{O}(MN\log N + NM\log M)$.
The inflation required by naive resampling methods then become yet more costly.

In this paper, we describe a variant of trigonometric interpolation, the \textit{LMN} method, that implements an optimized resampling of discrete waveforms in the frequency domain by avoiding the inflation that comes with naive algorithms consisting of an integer upsample factor followed by an integer downsample factor.
The quality and computational expense of this method are compared to others that are used in the field.

This work is motivated by a technical task required in the processing of data from detectors used by the Deep Underground Neutrino Experiment (DUNE)~\cite{DUNE:2020lwj}, a next-generation, long-baseline neutrino oscillation experiment.
DUNE’s primary physics objectives include determining the neutrino mass hierarchy, search for leptonic CP violation, and search for physics beyond the standard model of particle physics.
DUNE's primary detector technology is the liquid argon time projection chamber~\cite{Rubbia:1977zz,Chen:1976pp,Willis:1974gi} (LArTPC).
To understand the signals from these detectors, their operation is summarized.
When a charged particle traverses the liquid argon in a LArTPC detector, ionization electrons are produced along the particle's trajectory.
Under the influence of an external electric field, the ionization electrons drift at a constant velocity toward a stack of three parallel ``anode planes''.
Each anode plane is comprised of an array of parallel electrodes spanning the width of their plane.
The ionized electrons induce a negative electric current as they approach an electrode and a positive current as they drift past.
This induced current is amplified, filtered, digitized and sampled and the resulting discrete waveform is recorded.
Given the location of each electrode, these sampled waveforms represent a tomographic measurement of the distribution of drifting electrons in two dimensions: transversely across the electrode array and longitudinally along the electron drift direction.
No electron position information is available in the dimension along the length of the electrodes.
The anode planes are oriented so that their arrays of electrodes are at mutually unique angles.
Given a collection of these two-dimensional tomographic views from all anode planes, a three-dimensional image of the ionized electrons can be constructed.
From that image, it is possible to reconstruct information about the neutrino-argon interaction.
Finally, neutrino properties can be deduced from the ensemble of many such reconstructed interactions.
Lengthy and complex software developed by DUNE and the larger LArTPC community is applied to theses recorded waveforms to perform this extraction of physics information.

DUNE's far detector underground cavern provides for four independent cryostats, each containing approximately 17.5 kt of liquid argon. 
Currently, two far detector module designs are in the process of being implemented.
Each is based on a variant of common LArTPC design principles.
The modules are known by the direction of their electron drift: the first far-detector module FD1-HD~\cite{DUNE:2020txw}  employs a horizontal drift and the second FD2-VD~\cite{DUNE:2023nqi} drifts in the vertical direction.
They also utilize different anode plane and readout technologies.
In particular, the FD2-VD design divides its liquid argon volume into two regions and 
supplies anode plane electrodes as traces on large printed circuit boards
installed near the top and bottom of the cryostat.
Supported by the international DUNE collaboration, the design specifies different types of readout electronics for top and bottom volumes. 
The top electronics are based on the design developed for the dual-phase liquid argon detector technology~\cite{DUNE:2018mlo}.
The bottom electronics design~\cite{Radeka:2011zz} is in common with that used for the FD1-HD~\cite{DUNE:2020txw} module.
It is chosen to minimize the overall detector capacitance, which in turn increases the signal-to-noise ratio. 
This choice is driven by the necessarily long cable paths from the bottom anode planes to the cryostat roof. 
One consequence of employing different readout electronics technologies in FD2-VD is that a variety of sampling periods are employed. 
Specifically, the top electronics employs analog-digital converters (ADC) with a sampling period of \qty{500}{ns} while the bottom ADCs sample at \qty{512}{ns}. 
In addition, the FD1-HD will sample at \qty{512}{ns}.
Prior LArTPC experiments (MicroBooNE~\cite{MicroBooNE:2016pwy}, ProtoDUNE-SP~\cite{DUNE:2021hwx}, SBND) for which most of the existing data processing software was developed sample at \qty{500}{ns}.  

The readout electronics apply low-pass filters to their input analog signals in order to satisfy the Nyquist-Shannon~\cite{shannon1949communication} condition in the subsequent digitization stage.
Any sampling period that places the Nyquist frequency comfortably above the signal bandwidth may be chosen so that the information in the signal will be faithfully captured.
While the exact choice of sampling period is not critical, employing multiple sampling periods across the detectors leads to a technical complication for the data-processing software.
It is desirable to resample the ensemble of waveforms across the variety of detectors to produce waveforms with a common sampling period prior to applying the large software suite.
If the resampling assures the Nyquist frequency remains above the signal bandwidth and it does not otherwise distort the waveforms then this resampling will preserve all information relevant to the underlying physics.

This paper is organized as the following.
In \autoref{sec:sp}, the signal processing procedure applied to LArTPC data is briefly reviewed as it comprises the first stage of data processing and it places strong constraints on the quality of any resampling method.
The LMN resampling method is then described in \autoref{sec:method}. 
In \autoref{sec:cinoare}, the performance of the LMN method in terms of accuracy and computational cost is compared with some popular interpolation methods.
The application of the LMN method to realistic LArTPC data is shown in \autoref{sec:dune} using simulated data. 
Finally, a summary and discussions can be found in \autoref{sec:Summ}.

\section{LArTPC signal processing}\label{sec:sp}

When charged particles traverse the liquid argon medium, electrons are ionized along the particle trajectory and drift toward the anode planes under the influence of the external electric field. 
While drifting, the distribution of ionization electrons undergoes diffusion and absorption effects~\cite{Li:2015rqa}.
The TPC signal is then formed with i) the distribution of ionized electrons arriving at the anode plane, ii) the field response describing the resulting induced current on the anode electrodes~\cite{ramo1939currents} (see Ref.~\cite{Martynenko:2023bpe} for a recent example of calculating field response in three dimensions), iii) the electronics response, which amplifies, shapes and converts current to a voltage signal that is band limited, and iv) the ADC that samples the continuous voltage signal to produce discrete-time waveforms.

The goal of the signal processing procedure is to effectively invert this signal formation process.
It reconstructs a measure of the distribution of ionization electrons arriving at the anode electrodes given the recorded waveforms. 
The kernel of the procedure deconvolves a model of the overall detector response, which includes field and electronics responses, from the waveforms.
A technique to perform this deconvolution as a per-waveform filter that is applied in the frequency domain via discrete Fourier transform (DFT) was introduced to the LArTPC field in Ref.~\cite{Baller:2017ugz}.
It relied on a one-dimensional (1D) model that described how the drifting electrons induce current in an electrode as a function of time.
This was later improved by a new deconvolution that extending the model two dimensions (2D) by adding variation as a function of distance transverse to an electrode.
This new 2D method was first applied in the MicroBooNE experiment~\cite{MicroBooNE:2018swd, MicroBooNE:2018vro} where it demonstrated improvements compared to the 1D method.
The 2D method has since been utilized in the ProtoDUNE detectors, (DUNE prototype detectors), where it also achieved good performance~\cite{DUNE:2020cqd}. 

For simplicity and to share the notation of~\autoref{sec:method}, we describe the 1D technique that performs a deconvolution over the time dimension.  The 2D deconvolution is conceptually a straight-forward extension.
The technique extracts a finite, sampled sequence $\hat{s}[n]$ that estimates the number of electrons $q(t)$ drifting past a given electrode  $s[n] \equiv s(t_n) \equiv \int_{t=t_n}^{t_n+\Delta t} q(t)dt,\ n\in[0,...,N-1]$ over a sample time period $[t_n,t_n+\Delta t]$ to produce a sequence of $N$ samples.
This signal $s[n]$ is not directly observable as it represents the true distribution of ionization electrons.
Instead, a detector provides a measure $m[n]$ of the signal after the signal has been altered by a detector response modeled in general as $r[n',n]$.  In addition, noise is inescapable and is modeled as $\eta[n]$ following the measurement in data~\cite{MicroBooNE:2018swd}.  In terms of these sampled sequences, the measure is modeled as

\begin{equation}\label{eq:measure}
  m[n] = \sum_{n'} r[n',n] \cdot s[n'] + \eta[n].
\end{equation}
The first term is a convolution of the detector response and the signal.  
For time-invariant detectors, the response may be replaced with $r[n',n] \to r[n'-n]$.  This form allows the discrete Fourier transform (DFT, defined below in~\autoref{eq:fwddft}) to be applied to rewrite~\autoref{eq:measure} as $M[k] = R[k]S[k] + \mathcal{N}[k]$ were $k$ enumerates frequency domain samples. 
Neglecting noise for the moment allows for an estimate of the signal to be formed that may be expressed in the frequency domain as,
\begin{equation}\label{eq:estimate}
  S[k] \approx \hat{S}[k] = \frac{M[k]}{R[k]}.
\end{equation}

Naively, the inverse DFT (defined below in~\autoref{eq:invdft}) may be applied to recover a time domain estimate $\hat{s}[i]$.
However, several aspects of real detectors coincide to make that final step impractical.
First, the low-pass filter that is applied by the detector electronics to the analog input is designed to vanish at high frequency in order to satisfy the Nyquist-Shannon condition for the subsequent sampling.
Next, the electrodes comprising the initial anode planes do not collect drifting electrons.
This means their net induced current is zero and their response is bipolar in time resulting in $R[k\to 0] \to 0$.
Finally, real-world noise can not be neglected.
The noise to which LArTPC detectors are subjected fills a broad-band spectrum $\mathcal{N}[k]$ that can extend up to and beyond the Nyquist frequency while some sources will introduce noise in the circuit downstream of the band-limiting low-pass filters. 
Thus, the presence of noise in the measure $M[k]$ and a response $R[k]$ that is necessarily small or vanishing leads to~\autoref{eq:estimate} producing a naive signal estimate that is artificially amplified at high frequency and, for the initial anode planes, will diverge at low frequency.

To combat this practical problem, digital filtering is applied to the measure in both the frequency and time domains.
The time domain filter\footnote{The 2D technique also filters in the transverse or ``channel'' domain.} is introduced by rewriting \autoref{eq:estimate} as
\begin{equation}\label{eq:filter}
  S[k] \approx \hat{S}[k] = \frac{M[k]}{R[k]} F[k].
\end{equation}
From this result, an estimate $\hat{s}[n]$ in the time domain may be constructed by applying the inverse DFT.
In fact two such estimates are found, each with their own filter.
The first is constructed with a ``Wiener-inspired'' filter that maximizes signal-to-noise.
The second uses a (truncated) Gaussian-shaped filter to give an unbiased measure of the signal.

The time-domain digital filtering consists of identifying ``signal regions-of-interest'' (ROIs).  Each ROI selects a fragment of the signal estimate after the Wiener-inspired filtered has been applied and that is expected to have a high signal-to-noise ratio.
The algorithms that identify the signal-ROIs are complex heuristics created by a development process that considered data from real LArTPC detectors such as MicroBooNE~\cite{MicroBooNE:2018swd,Yu:2020wxu}.
These ROIs are then applied to the Gaussian-filtered estimate and a new signal baseline is calculated for each region based on values at either ends of the ROI. 
This process effectively combats the problematic combination of noise with the small or vanishing portions of the response spectrum.

\section{LMN resampling method}~\label{sec:method}

We review established resampling methods on which the LMN method is based. 
We then describe the LMN method, which is essentially identical to the basic method but adds a constructed, special-case optimization.
Finally, a discussion of normalization, information loss and mitigation of artifacts is given.

\subsection{Resampling basics}
\label{sec:resbasics}

A continuous signal \(g(t)\) may be sampled at times \(t_n = n\cdot T_s, n\in [0,...,N_s-1]\) with period \(T_s\) to produce a time-domain sequence \(\vec{g}_s \equiv \{g_s[n]\},\ g_s[n] \equiv g(t_n)\).
From this sequence, another sequence
\(\vec{g}_r \equiv \{g_r[n]\}, n\in [0,...,N_r-1]\) may be constructed that corresponds to a \emph{resampling} of \(g(t)\) with a sampling period \(T_r \ne T_s\).
This correspondence may be exact or approximate 
depending on the resampling method, the bandwidth of the signal and whether an overall \emph{upsampling} (\(T_r < T_s\)) or \emph{downsampling} (\(T_r > T_s\)) is performed.

One general resampling method known as \emph{fractional resampling}~\cite{schafer:1973} decomposes the resampling procedure into two stages.  First, the original sequence \(\vec{g}_s\) is upsampled by an integer factor \(L\) to produce an intermediate sequence \(\vec{g}_u\) with a larger number of samples \(N_u = LN_s\).
Second, this intermediate sequence is downsampled by an integer factor \(M\) to produce the final sequence \(\vec{g}_r\) with a number of samples \(N_r = \frac{N_u}{M}\).
The \emph{simple fraction} \(\mrat\equiv\frac{L}{M} = \frac{N_r}{N_s}\) characterizes the fractional resampling.

If the duration of the original signal is unchanged in the resampling then \(N_s T_s = N_r T_r\) and thus \(\mrat = \frac{T_s}{T_r}\).  However, if \(\frac{T_s}{T_r}\) is irrational then fractional resampling can not be applied exactly.  In such cases, a simple fraction \(\mrat \approx \frac{T_s}{T_r}\) may be chosen to approximate the desired resampling ratio.  The choice can be arbitrarily precise at the cost of potentially large values of \(L\) and \(M\).

\subsection{Discrete Fourier transform}
\label{dft}

The resampling used here is based on a subclass of these integer fractional methods that operate on the Fourier representation of the sampled signal~\cite{fraser:1989}.
The discrete Fourier transform (DFT) is applied to the time-domain sequence \(\vec{g}\) of size \(N\) to produce the frequency-domain sequence \(\vec{G} \equiv \{G[k]\} = DFT(\vec{g})\) with

\begin{equation}
\label{eq:fwddft}
G[k] = \sum_{n=0}^{N-1}g[n] \cdot e^{-i2\pi \frac{nk}{N}},\ k \in \mathbb{I}.
\end{equation}
This sequence is infinite and periodic, \(G[k] = G[k+N]\) over size \(N\), and has two points of Hermitian symmetry which are important for the resampling.
The first symmetry is around the ``zero frequency'' sample \(G[0]\) so that \(G[k] = G^*[-k]\).
The second is around the Nyquist frequency and is expressed differently if the number of samples is even or odd as

\begin{equation}
\label{eq:nyqsym}
G[H-k] =
\begin{cases}
G^*[H+k+2], &  N \in \mathcal{N}_{even} \\
G^*[H+k+1], &  N \in \mathcal{N}_{odd}
\end{cases}
\end{equation}
Here, $H$ gives the size the ``half spectrum'' that consists of the sub-sequence of frequency-domain samples that are not Hermitian-symmetric with themselves.  It is given by
\begin{equation}
\label{eq:aitch}
H =
\begin{cases}
\frac{N}{2}-1 ,& N \in \mathcal{N}_{even} \\
\frac{N-1}{2} ,& N \in \mathcal{N}_{odd}
\end{cases}
\end{equation}
For \(N \in \mathbb{N}_{even}\), the sample \(G[H+1]\) is real and called the ``Nyquist bin'' as the Nyquist frequency lands at the center of its frequency band.  For \(N\in \mathbb{N}_{odd}\), the Nyquist frequency is at the lower edge of the sample \(G[H+1]\).

There are two common ways to represent a finite subsequence of \(\vec{G}\) of size \(N\).
First is the ``Nyquist-centered'' range where the subset of samples \(\{G[1],...,G[H]\}\) are Hermitian-reflected about the Nyquist frequency starting with \(G[H+1] = G^*[H]\) for \(N \in \mathbb{N}_{odd}\) or with \(G[H+2] = G^*[H]\) for \(N\in \mathbb{N}_{even}\).
The second is the ``zero-centered'' range where
the same subsequence of size \(H\) is instead Hermitian-reflected about \(G[0]\).
Influenced by this second choice, the reflected subsequence is often said to consist of the ``negative frequency'' samples.

The inverse DFT may be applied to \(\vec{G}\) to recover the original time-domain sequence \(\vec{g}\) as
\begin{equation}
\label{eq:invdft}
g[n] = \frac{1}{N}\sum_{k=0}^{N-1}G[k] \cdot e^{i2\pi \frac{nk}{N}},\ n \in \mathbb{I}.
\end{equation}
The result of the inverse DFT is also an infinite, periodic sequence.  Note, the form of the normalization term \(\frac{1}{N}\) shown here is the convention followed by most software providing the DFT.  Additional discussion of normalization is in Sec.~\ref{sec:norm}.

\subsection{Resampling in the Fourier domain}
\label{sec:dftres}

An upsampling from \(N_s\) to \(N_u = LN_s\) can be accomplished in the frequency-domain by 
inserting \((L-1)N\) samples at both ends of the zero-centered range (or equivalently  in the middle of the Nyquist-centered range) of \(\vec{G}_s\) to produce an intermediate spectrum sequence \(\vec{G}_u\) with samples giving,

\begin{equation}
\label{eq:fdus}
G_u[k] =
\begin{cases}
G_s[k], & k \in [-H_s, ..., 0, ..., H_s] \\
0     , & \text{otherwise}
\end{cases}
\end{equation}
Here, \(H_s\) represents the ``half spectrum'' size for the original sampling as defined in Eq.~\eqref{eq:aitch} and it is assumed that \(N_s \in \mathbb{N}_{odd}\).
When \(N_s \in \mathbb{N}_{even}\), Eq.~\eqref{eq:fdus} applies with the addition that the Nyquist bin at \(G_s[H+1]\) requires special treatment which is described in Sec.~\ref{sec:nyqbin}.

If \(\frac{N_u}{M} \not\in \mathbb{N}\) then the final \(M\)-wise downsampling 
must be performed in the time domain after applying the inverse DFT of (large) size \(N_u = LN_s\) to \(\vec{G}_u\).
This allows any remainder sequence of interval domain samples of size \(N_u\ \mathrm{mod}\ M\) to be ignored and corresponds to an effective change in the duration of the resampled signal compared to the original.

In the special case that \(N_u/M \in \mathbb{N}\), the downsampling may be performed in the Fourier domain by removing  \(\frac{N_u}{M}(M-1)\) samples from the middle of the Nyquist-centered range to produce \(\vec{G}_r\) of size \(N_r\) with terms (expressed zero-centered),
\begin{equation}
\label{eq:fdds}
G_r[k] = G_u[k], k \in [-H_r, ..., 0, ..., H_r].
\end{equation}
Here again, \(H_r\) gives the ``half spectrum'' size as defined in Eq.~\eqref{eq:aitch} for the resampled sequence. Eq.~\eqref{eq:fdds} is strictly for \(N_r \in \mathbb{N}_{odd}\) and treatment of the Nyquist bin for the case \(N_r \in \mathbb{N}_{even}\) is put aside until Sec.~\ref{sec:nyqbin}.

Obviously, in this special case, the upsampling and downsampling stages may be combined so that the net number of frequency-domain samples is changed by \(\Delta N = N_r - N_s\).  When \(\Delta N > 0\) a net upsampling is performed by inserting this number of zero-valued samples similar to Eq.~\eqref{eq:fdus}.  
Conversely, \(\Delta N < 0\) indicates a net downsampling is performed with (the absolute value of) this number of samples removed as in Eq.~\eqref{eq:fdds}.
To finish, the intermediate spectrum \(\vec{G}_r\) of size \(N_r\) is formed and the inverse DFT is applied to produce a resampled time-domain sequence \(\vec{g}_r\).  In this special case, the size of the forward DFT is $N_s$ and the size of the inverse DFT is $N_r$, both of which are typically many factors smaller than the inflated $N_u$.  It is seeking this special case that leads to the LMN optimization.

\subsection{Special-case optimization}
\label{sec:specialcase}

To give it a name, creating this special case or simply exploiting it when it is accidentally met is called here the \textit{LMN method}.
The special case \(\frac{NL}{M}\in \mathbb{N}\)  can be met for any given \(T_s\) and \(T_r\) that meet a rationality condition described below and when there is freedom to choose a value of \(N_s\) even if it may differ from the size of a given input signal.
It is possible to generate a series of values for \(N_s\) that satisfy \(N_sL/M \in \mathbb{N}\).  
Constructing the special case begins by considering that the resampling will shift the sampling frequency by an amount

\begin{equation}
    \label{eq:nfshift}
    \Delta F = \frac{1}{T_r} - \frac{1}{T_s} = \frac{T_s-T_r}{T_sT_r}.
\end{equation}
Next, this shift must span an integer number \(\delta N \in \mathbb{N}\) of original frequency-domain samples and thus must cover a frequency range

\begin{equation}
    \label{eq:ibshift}
    \delta F = \frac{\delta N}{N_sT_s}.
\end{equation}
Equating \(\delta F = \Delta F\) results in the relation

\begin{equation}
    \label{eq:nfibshift}
    N_s = \frac{\delta N \cdot T_r}{T_s - T_r} \in \mathbb{N}.
\end{equation}
The smallest value of \(|\delta N|\) that satisfies Eq.~\eqref{eq:nfibshift} results in a minimal sequence size \(N_s = N_{min}\) for which the special case applies.  Any multiple of \(N_{min}\) will also satisfy the special case.
If found, we may select the multiple of \(N_{min}\) that best approximates the size of a given input signal.

The satisfactory value for \(\delta N\) may be found as
\begin{equation}
    \label{eq:nsolve}
    \delta N = \frac{T_s - T_r}{\gcd(T_r,\ T_s - T_r)}.
\end{equation}
Here, \(\gcd(\cdot,\cdot)\) returns the greatest common divisor of its arguments.
As the arguments are in general real valued, this function may apply the Euclidean algorithm to calculate its result.
Computer floating-point round-off error will require the algorithm to test for convergence against a small, non-zero error parameter.
Note, a lack of convergence within this error indicates that the ratio of its arguments is not rational.
When this \textit{LMN rationality condition} is not met, this form of fractional resampling can not be applied exactly.  
This is similar to the rationality condition of the integer fraction method on which LMN is based.

The choice of the special case size $N_s$ of course determines the size of the forward DFT and (along with the chosen resampling ratio \rat) the size of the inverse DFT. 
The freedom to select a multiple of $\delta N$ may be exploited for a secondary optimization.
The speed of the DFT strongly depends on the size of prime factors that make up the length of the sequence being transformed~\cite{Good:1958}.
A DFT size with large prime factors requires substantially more processing than nearby larger sizes.
Thus, all else being equal, an application of the LMN method may find that an otherwise unnecessary extension  of the original interval domain sequence by an additional $\delta N$ may lead to an overall faster resampling.
The effects of this dependence on the sizes of the prime factors of a sequence length can be seen in the performance results presented in~\autoref{sec:cinoare}.

\subsection{Treatment of Nyquist bins}
\label{sec:nyqbin}

In general, when \(N \in \mathbb{N}_{even}\) the sequence \(\vec{G}\) possesses a Nyquist bin sample.  When resampling in the Fourier domain, this sample requires special treatment in two cases. The first case~\cite{fraser:1989} is when a net upsampling from size \(N_s \in \mathbb{N}_{even}\) is performed.
Half of the Nyquist bin sample value $G_{s,nyquist} = G_s[H_s+1]$ is transferred to each of the two equivalent samples in the resampled spectrum,
\(G_r[H_s + 1] = \frac{1}{2}G_{s,nyquist}\)  and \(G_r[-H_s - 1] = \frac{1}{2}G_{s,nyquist}\).  Note, neither of these two samples in the upsampled spectrum \(\vec{G}_r\) are themselves a Nyquist bin.
The second and opposite case is when a net downsampling to size \(N_r \in \mathbb{N}_{even}\) is performed.
The Nyquist bin in the resampled spectrum must be set to the absolute value of either of the two corresponding samples from the original spectrum, for example as \(G_r[H_r + 1] = |G_s[H_r + 1]|\).

\subsection{Normalization}
\label{sec:norm}

A ``round trip'' of a sequence through the forward and inverse DFT requires an overall normalization of \(\frac{1}{N}\) shown in Eq.~\eqref{eq:invdft}.
We follow here the dominant convention to apply this normalization fully as part of the inverse DFT.
Other conventions are possible as long as a total of \(\frac{1}{N}\) normalization is applied in the round trip.
For example the symmetric normalization places $\frac{1}{\sqrt{N}}$ on both forward and inverse DFT.
All such conventions produce the same result as long as \(N\) is unchanged throughout the round trip.

However, resampling in the Fourier domain necessarily performs the forward DFT with size \(N_s\) and the inverse DFT with size \(N_r \ne N_s\).
Care is needed to select a normalization convention that preserves a desired invariant.
The choice of invariant may depend on how an application interprets the quantity measured by signal samples.

For each of three interpretations considered below, the nominal \(\frac{1}{N}\) applied as part of the inverse DFT is taken for granted and an additional normalization of the form \(g_r[n] \to g'_r[n] = Ag_r[n]\) is provided.  It is also assumed that only zero-value frequency samples are inserted or removed.  Later we consider the case where this assumption is not held.

\begin{itemize}
\item {\bf Interpolation:}
  An interval-domain sample gives a measure of the instantaneous value of the underlying continuous signal.
  This interpretation requires resampling to preserve the local amplitude of the interval-domain samples and thus act as an \textit{interpolation}.
  The nominal DFT convention applies a normalization of \(\frac{1}{N_r}\) and so the signal amplitude will be reduced in an upsampling and increased in a downsampling compared the a direct resampling of the continuous signal.
  Thus, an additional normalization of \(A_0 = \frac{N_r}{N_s}\) must be applied to the resampled sequence for correct interpolation.
  Alternatively, applying this normalization is equivalent to applying the basic DFT normalization fully as part of the \textit{forward} DFT instead of following the accepted convention to apply it fully as part of the inverse DFT.

\item {\bf Integral: } An interval-domain sample represents the integral of the underlying continuous signal over the sample period.
This interpretation must preserve the sum of the time-domain sequence samples through the resampling process as expressed by
\begin{equation}
\label{sumcons}
\sum_{n=0}^{N_s-1} g_s[n] = A_1A_0\sum_{n=0}^{N_r-1} g_r[n].
\end{equation}
If we were to apply only the interpolation normalization \(A_0\) it would be equivalent to integrating each of \(N_r\) points over the original sample period \(T_s\) while integration over the resampled period \(T_r\) is proper to preserve the equality between sampled and resampled sums.  Thus an additional normalization of \(A_1 = \frac{T_r}{T_s} = \frac{N_s}{N_r}\) must be applied in addition to the interpolation normalization.
These two normalization terms cancel \(A_1A_0 = 1\) and thus the nominal DFT normalization convention is directly appropriate for the integral interpretation.

\item {\bf Energy: } An interval-domain sample gives a measure of a probability amplitude, field or other quantity where the sum-of-squares must be conserved through the resampling.
  This sum is conserved through the forward DFT as 
expressed by Parseval's theorem which equates an energy measure in the time domain \(E_t\) to an energy measure in the frequency domain \(E_f\).  In terms of the original sampling, this may be written as
\begin{equation}
\label{parseval}
E_{t,s} \equiv \sum_{n=0}^{N_s-1} \left|g_s[n]\right|^2 = \frac{1}{N_s}\sum_{k=0}^{N_s-1} \left|G_s[k]\right|^2 \equiv E_{f,s}.
\end{equation}
The energy after resampling may be written as
\begin{equation}
\label{parsevalf}
E_{f,r} \equiv \frac{1}{N_r}\sum_{k=0}^{N_r-1} \left|G_r[k]\right|^2 = \frac{1}{N_r}\sum_{k=0}^{N_s-1} \left|G_s[k]\right|^2 = \frac{N_s}{N_r}E_{f,s}.
\end{equation}
The effective replacement \(r \to s\) in the sum of right hand term is allowed as we have assumed only zero-amplitude frequency samples have been added or removed in the resampling.  Thus, in order for Parseval energy to be conserved through the resampling, an additional normalization \(A_2 = \sqrt\frac{N_r}{N_s}\) must be multiplied to the resampled time-domain sequence, \(g_r[n] \to g'_r[n] = A_2g_r[n]\).
\end{itemize}

In the above discussion, it is assumed that only zero-amplitude frequency-domain samples are inserted or removed in the resampling.  In the case of upsampling, this may assured by construction.  In the case of downsampling it is only assured if the original signal is band-limited to be below the new, smaller Nyquist frequency.  When this band limit is not assured, a downsampling effectively applies a perfect low-pass filter with a cut-off at the resampled Nyquist frequency \(\frac{1}{2T_r}\).  The removal of the non-zero samples represents actual loss of signal information.

\subsection{Potential artifacts}

The method just described, and indeed any based on a round-trip via DFT, assumes the signals are periodic and band-limited.  Real-world signals are typically neither.  Resampling such signals can induce artifacts.


The original sampling produces a sequence over a finite interval window.  It is typically inescapable that the end of that sequence will not ``wrap around'' to form a smooth connection to the beginning of the sequence.  Even in a special case where an underlying signal is constructed to be periodic it typically includes aperiodic noise.

In the frequency domain, another source of artifacts can arise.  While it is typical to band-limit signals with a low-pass filter prior to the original sampling, real world low-pass filters merely attenuate and do not fully remove power above their ``cut-off'' frequency.  Another source of high-frequency content is noise that is picked up by  electronics circuit components that exist after the low-pass filter.  Insertion or removal of frequency samples is effectively equivalent to the application of a sharp low-pass filter.  This results in some amount of ringing in the interval domain after the inverse DFT is applied.


The LMN method provides ways to mitigate the endpoint discontinuities of non-periodic signals. 
To achieve the special case $N_s$ size, the original signal must either be truncated or extended in the interval domain. 
When extended, there is freedom to choose the values applied when padding the signal to size $N_s$. 
Simple schemes can be employed to allow the endpoints of the signal to be smoothly, periodically connected. 
For example, padding can be filled with an interval-domain linear interpolation between last and first sampled values of the original signal. 
Depending on the degree of discontinuity, the sharp ``kinks'' this can create may still produce ringing. 
If significant, more sophisticated padding models can be applied to form a smoother connection.
If the application allows, these artifacts can also be mitigated by exploiting the intermediate frequency-domain representation. 
There, a low-pass filter spectrum can be multiplied.
Both of these forms of mitigation require minimal additional computational expense.

\section{Performance comparison with selected methods}\label{sec:cinoare}
The performance of a resampling algorithm can be evaluated in different ways and it will vary subject to the shape of the considered signal as well as the resampling ratio. 
Therefore, it is important to evaluate various performance metrics and to do so for a range of signal frequencies, noise levels, and sampling ratios. 
In this section, the performance of the LMN method in a waveform resampling task will be compared with selected interpolation methods including: i) linear interpolation, ii) polyphase filtering, and iii) cardinal cubic B-spline.

\subsection{Considered signals}
\begin{figure}
  \includegraphics{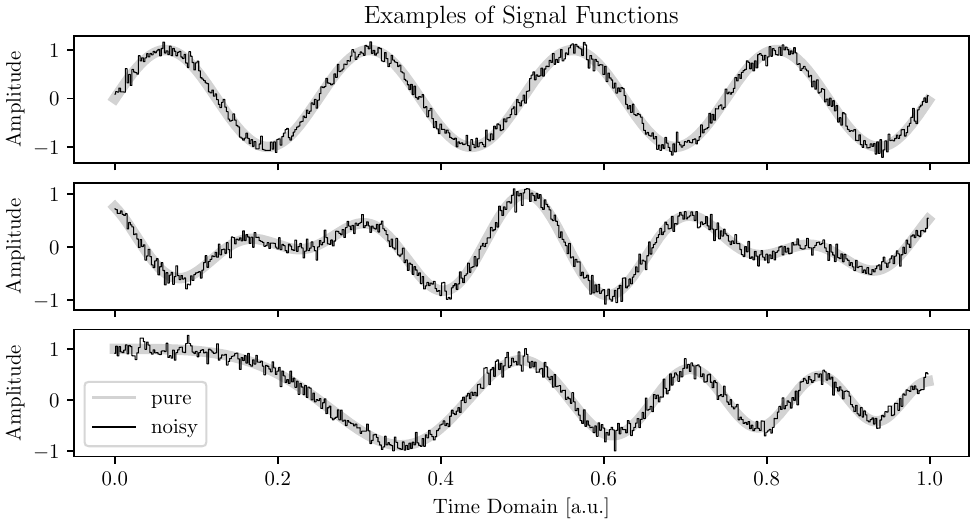}
  \caption{Examples of generated input signals. The following underlying true functions were used from top to bottom: \textit{single freq}, \textit{multi freq}, \textit{gauss pulse}. All three were generated with a frequency parameter of $c_f=4$ and a noise amplitude of $c_n=0.1$.}
  \label{fig:perf:signals}
\end{figure}
The measure introduced in \autoref{eq:measure} is idealized by a signal generated from an analytic function $-1\le f \le 1$ representing the convolution $s * r$ and a white-noise term $c_n\eta$ formed by scaling samples drawn from the normal distribution.
This idealized measure is then sampled at period $T_s \equiv \frac{1}{N_s}$ to produce an initial sampled sequence $g[n] = f[n] + c_n \cdot \eta[n],\ n\in[0,...,N_s-1]$.

\begin{flalign}
 & & f(t) &= \sin(c_f \cdot t) & & \text{\textit{single freq}}\label{eq:single-freq}  \\
 & & f(t) &= \frac{1}{2} (\cos(c_f \cdot t) + \cos(1.4 \cdot c_f \cdot t + \frac{\pi}{3})) & & \text{\textit{multi freq}}\label{eq:multi-freq} \\
 & & f(t) &= \exp(-t^2) \cdot \cos(c_f \cdot t^2) & & \text{\textit{gauss pulse}}\label{eq:gauss-pulse}
\end{flalign}

Here, $c_f$ is a frequency parameter that allows to generate a similar signal with a different frequency. Examples of such signals can be seen in \autoref{fig:perf:signals}.  These waveforms have duration of 1.0 unit of time and $N=500$ samples and thus a sample period of $0.002$ units of time and a Nyquist frequency of 250 cycles per unit of time.

\subsection{Performance metrics}
\begin{figure}
  \includegraphics{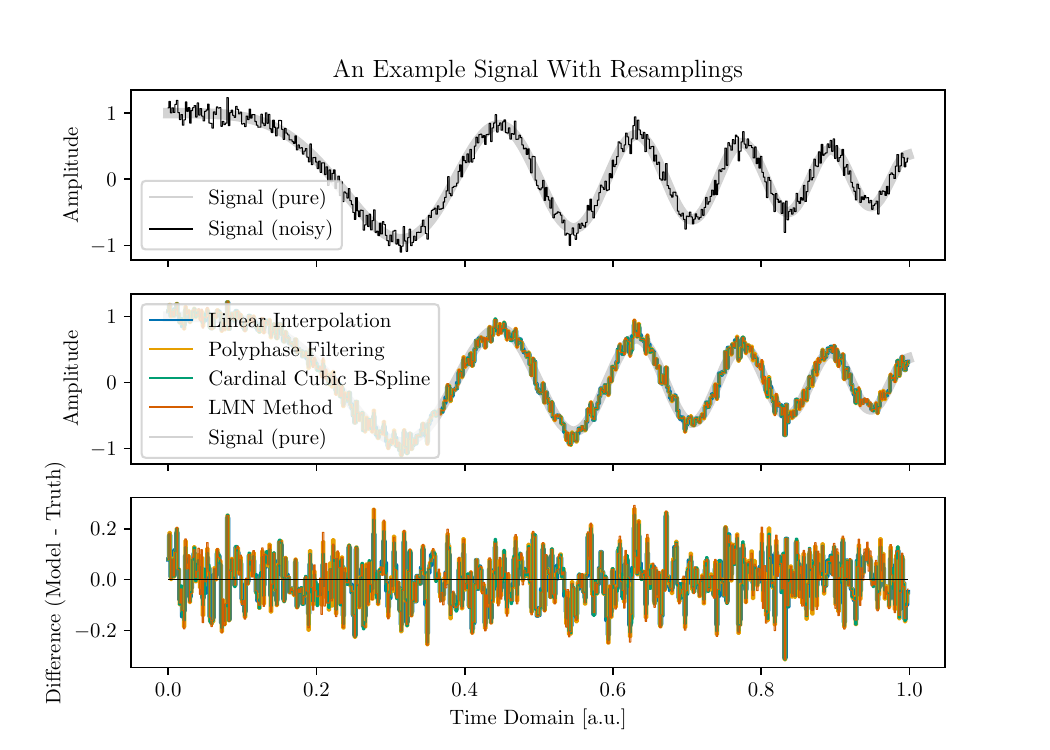}
  \caption{Comparison of true underlying signal with resampled signal. For each case, the signal was resampled from $N_s=500$ to $N_r=512$ points. The assumed signal here is \textit{gauss pulse} with a frequency parameter of $c_f=5$.}
    \label{fig:perf:triple}
\end{figure}
The initial sequences are resampled $g[n]\to g_r[n'],\ n\in[0,...,N_s-1],\ n'\in[0,...,N_r-1]$.
The top of~\autoref{fig:perf:triple} shows an example of a signal of the \textit{gauss pulse} type (\autoref{eq:gauss-pulse}) with one choice of frequency and noise parameters.
In the middle, the results of resampling this signal from $N_s=500$ to $N_r=512$ points while conserving total signal duration are shown for the considered resampling methods.
This is representative of upsampling a signal with sample period \qty{512}{ns} to one with \qty{500}{ns}.
The bottom panel shows the difference between the resampled discrete noisy signal (``Model'') and the ``pure'' noise-free signal function $f(t)$ evaluated at the resampled times (``Truth'').

\subsection{Scaling with signal frequency}
In order to evaluate the quality of a resampling across a wide spectrum of signal frequencies, sequences were generated with the \textit{single freq} function (\autoref{eq:single-freq}) for different values of the frequency parameter, $c_f$. Each sequence was then resampled from $N_s=500$ to $N_r=512$ points using the considered resampling methods.
To quantify how well the resampled signal matches the underlying true signal, the mean-squared error (MSE) statistic of~\autoref{eq:mse} is calculated.

\begin{equation}
  \label{eq:mse}
  \text{MSE} = \frac{1}{N_r}\sum_{i=0}^{N_r-1}{(f[n]-g_r[n])^2}.  
\end{equation}

\begin{figure}
  \includegraphics[width=\textwidth]{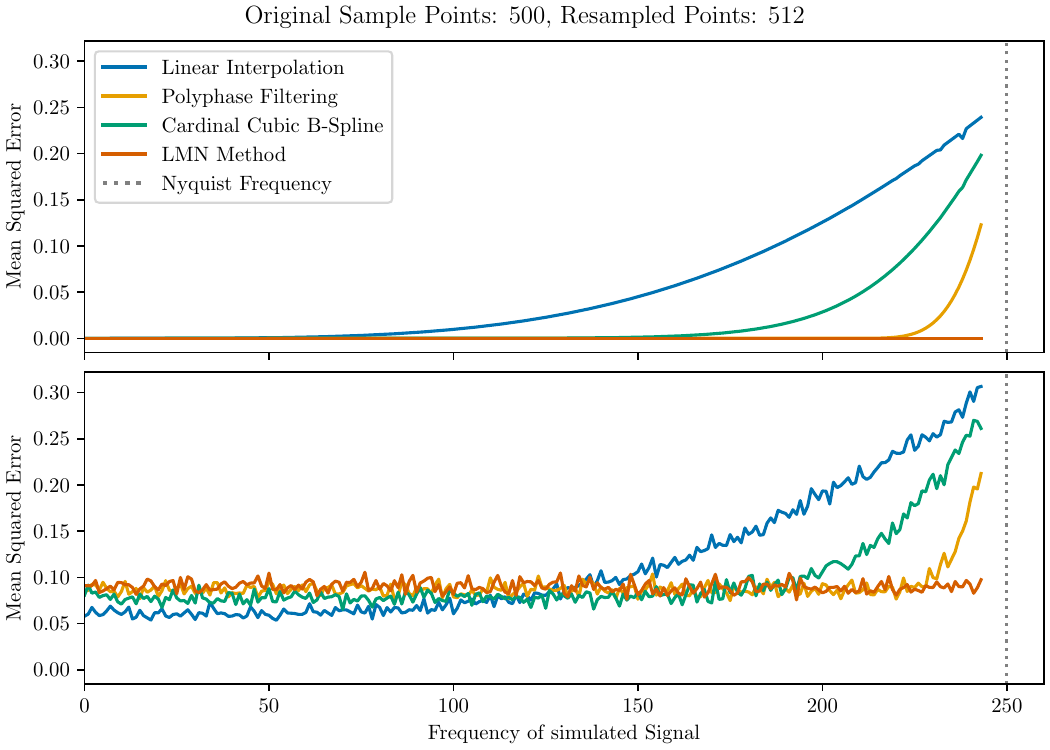}
  \caption{Mean squared error as a function of the frequency of the simulated signal. The \textit{single freq} signal was used with a noise amplitude of $c_n=0$ and $c_n=0.3$ in the top and bottom, respectively. For each case, the signal was upsampled from $N_s=500$ to $N_r=512$ points. 
  The results of the Naive Resampler are omitted since they are identical to those of the LMN method.
  }
  \label{fig:perf:freq_scaling}
\end{figure}

In \autoref{fig:perf:freq_scaling}, the MSE as a function of the signal frequency parameter $c_f$ is shown for the case of no noise (top panel) and a chosen non-zero value of $c_n$ (bottom panel). 
For the noiseless case, it can be seen that the LMN method results in mean squared errors very close to or identical to zero across all considered values for $c_f$. 
The other considered resampling methods exhibit higher MSE with increasing signal frequency parameter. 
In the case of a noisy signal, the MSE for signals with a small frequency parameter is lowest for the linear interpolation. 
This is not necessarily desired, as the noise level should in principle remain untouched by resampling.
This reduction of noise is the consequence of an implicit spectral distortion.

\begin{figure}
  \includegraphics[width=\textwidth]{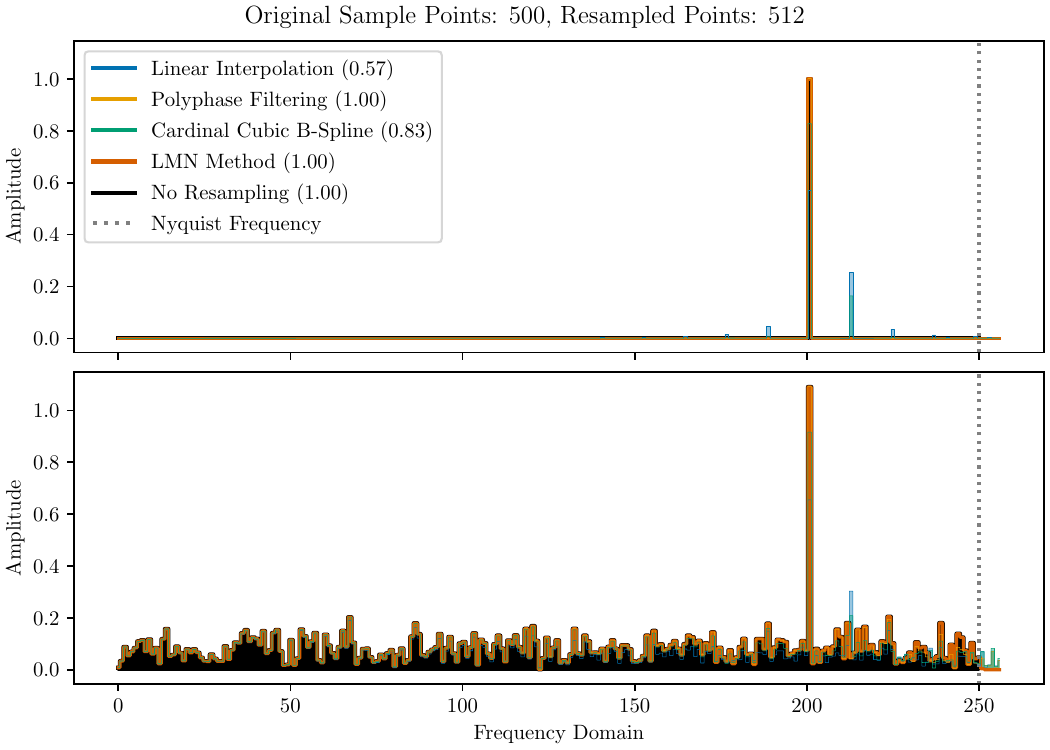}
  \caption{Frequency-domain spectra of original and resampled signals.  The \textit{single freq} underlying signal was used with a frequency parameter of $c_f=200$ and a noise amplitude of $c_n=0$ and $c_n=1$ in the top and bottom, respectively. For each case, the signal was upsampled from $N_s=500$ to $N_r=512$ points. The amplitude at the peak frequency is given in parentheses.  The original Nyquist frequency is marked with the dashed line.}
  \label{fig:perf:artifacts}
\end{figure}

In order to understand the inaccuracy of some of the resampling methods at high signal frequencies,
frequency-domain spectra of original and resampled sequences are show in in~\autoref{fig:perf:artifacts} for the case $c_f = 200$.
In the figure, the original signal spectrum is nearly identical to the resampled spectra from the LMN and polyphase filtering methods.
Resampling with the other methods, however, introduces artifacts below and above the signal frequency and show reduced power at the sole frequency of the true signal.

\subsection{Processing time scaling with sampling rate}
\begin{figure}
  \includegraphics[width=\textwidth]{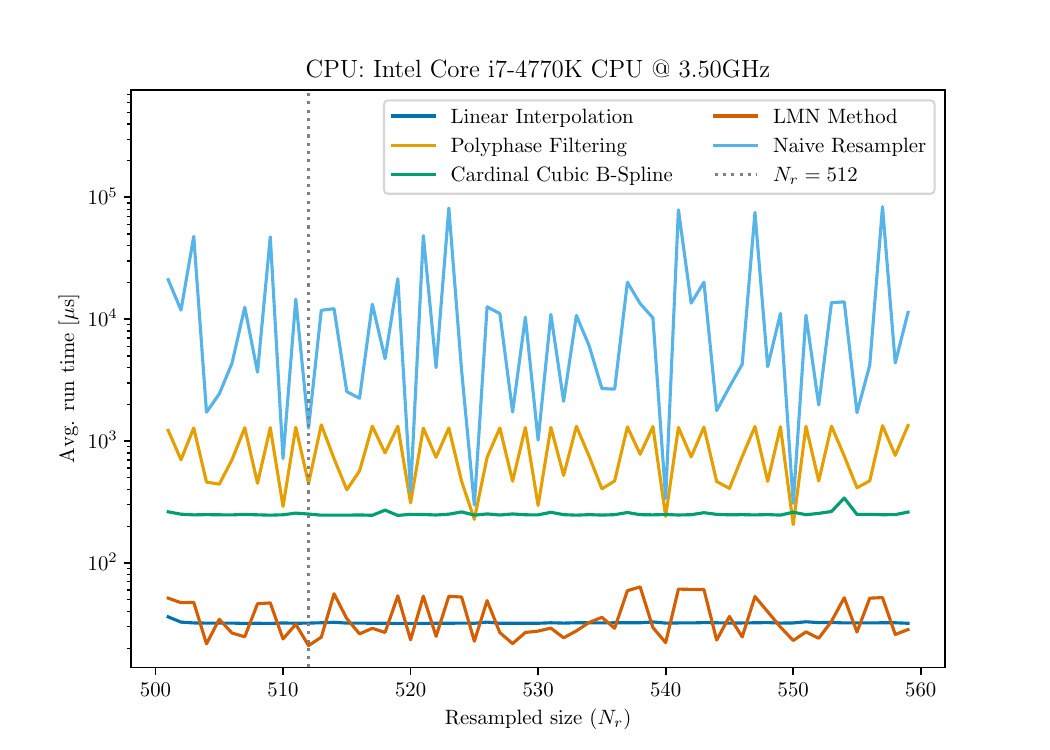}
  \caption{Average processing (run) time vs. resampled size for different methods.  See text for explanation of the Naive Resampler.}
  \label{fig:perf:freq_space}
\end{figure}

Besides being lossless, a good resampling method should also be computationally efficient.
As a way to evaluate this efficiency, the various resampling algorithms are deployed to resample the same noisy signal 100 times from which average processing times are determined.
The original signal with $N_s=500$ data points is resampled to sizes between 501 and 565 points.
The average processing time as a function of the resampled size is shown in 
\autoref{fig:perf:freq_space}.
There the result of the large inflation required for the Naive Resampler is reflected as large running times.
All methods utilizing DFT exhibit large variance in average processing times across the different resampled sizes, $N_r$.  As also mentioned in \autoref{sec:specialcase}, DFT processing time depends on not just the total size of a waveform but the exact set of prime factors of the waveform size.  A size that is factored by large primes requires more computation than similar sizes that factor into only small primes.
Depending on the precise number of resampled data points, either the LMN method or linear interpolation requires the least computation, whereas the naive approach is typically two orders of magnitude slower.

\section{LArTPC waveform resampling}\label{sec:dune}

The motivation for developing the LMN resampling method was to solve the practical and technical challenges of processing large amounts of data with differing sampling periods from the different DUNE LArTPC detectors.
Thus it is imperative to demonstrate the method on waveforms that are representative of what the LArTPC detectors produce.
This section illustrates the resampling of waveforms with period \qty{512}{ns} to those with period \qty{500}{ns}.  
The waveforms are produced by the current state-of-the-art detailed LArTPC signal and noise simulation algorithms from the Wire-Cell Toolkit~\cite{MicroBooNE:2018swd,wcsim}. The simulations of both signal and noise have been validated by data measurements~\cite{MicroBooNE:2018swd,DUNE:2020cqd}. 

\begin{figure}
  \includegraphics[page=1,width=0.9\textwidth]{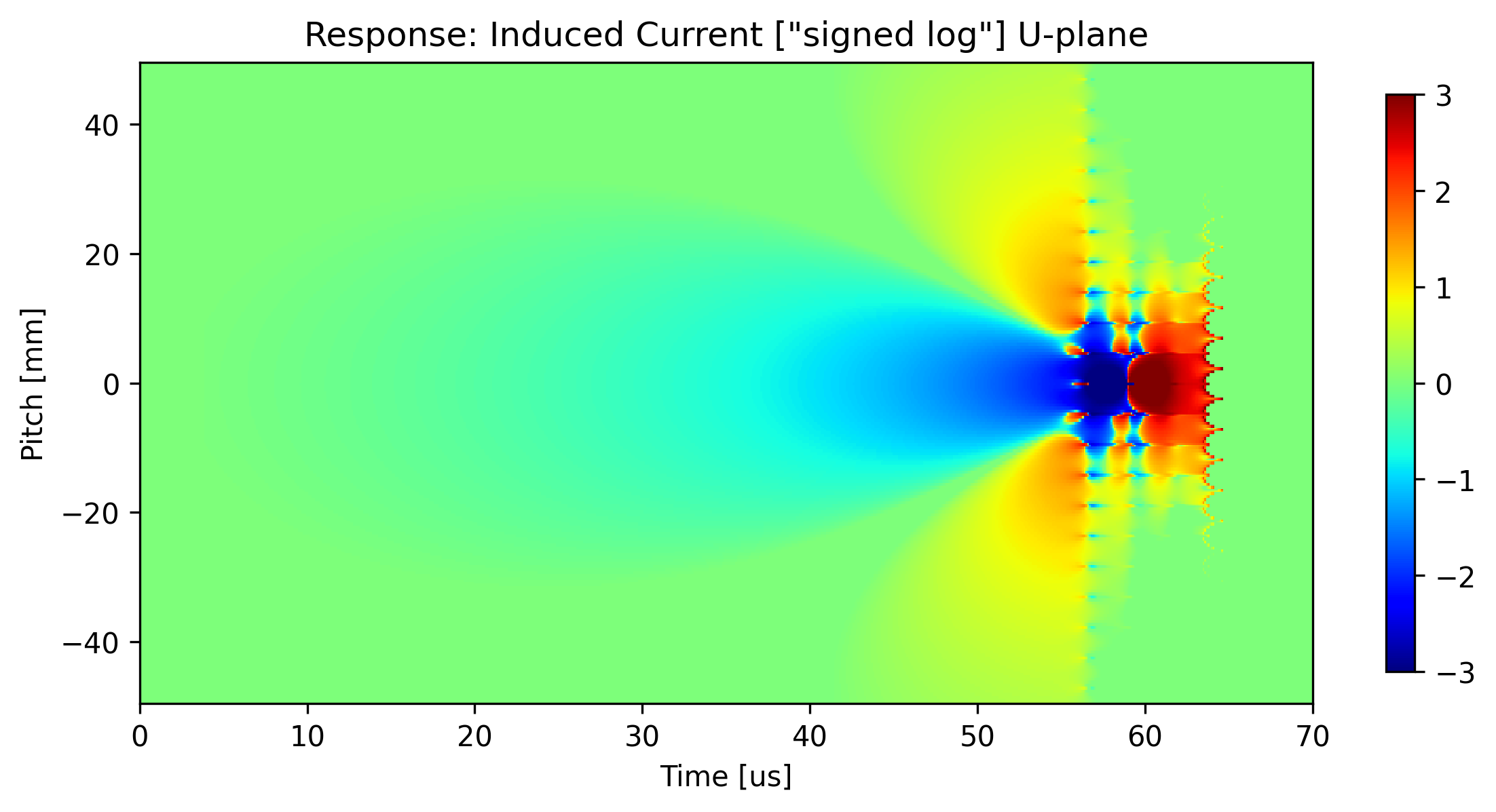}
  \caption{The induced electric current field responses for the first anode plane of the DUNE horizontal drift detector design. The ``pitch'' axis samples transverse to the nominal electron drift direction while the ``time'' axis samples at \qty{100}{ns} steps along the detailed drift path.  Color gives a logarithmic scale for the induced current at a given step.}
  \label{fig:frs}
\end{figure}

This simulation models the physical processes involved in electron drift~\cite{Li:2015rqa,Li_2022} through liquid argon, induced current in anode electrodes via field and electronics response, noise effects and ADC digitization as introduced in \autoref{sec:sp}.
The kernel of the simulation involves a convolution of the distribution of ionization electrons after they have drifted through the volume with the field and electronics responses. 
The field responses are calculated with a 2D model of the anode electrodes, the various applied electrical potentials and a model of electron mobility in liquid argon. 
An example of the resulting field responses for one anode plane, calculated with GARFIELD~\cite{Veenhof:1993hz}, is shown in~\autoref{fig:frs}. 
It is represented as a 2D array giving the amount of induced electric current in an electrode of interest at different electron drift path starting locations transverse to (``Pitch'') the electrode and along (``Time'') the nominal drift path. 
The longitudinal dimension is sampled with a period of \qty{100}{ns}. 
As part of the convolution kernel operation, the simulation will apply an integer downsampling. 
With \qty{100}{ns} responses, a downsampling factor of five achieves the target ADC sample period of \qty{500}{ns}.

The \qty{100}{ns} sampling can not be downsampled by an integer factor to the target ADC sample period of \qty{512}{ns} and so a new set of field responses are needed. 
As the field calculations are computationally costly, we elect here to apply LMN resampling to the results at the nominal \qty{100}{ns} to produce results sampled at \qty{64}{ns}. 
This allows the simulation to downsample by an integer factor of eight and achieve the desired \qty{512}{ns} ADC sampling period. 
This resampling of the field responses also provides an auxiliary test of LMN method.

\begin{figure}
  \includegraphics[width=0.9\textwidth,page=1]{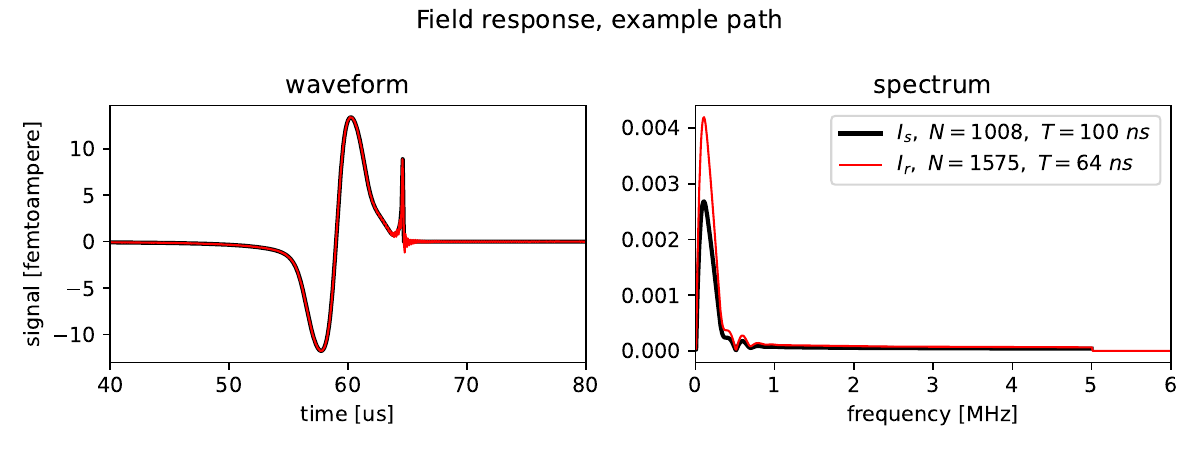}
  \caption{One induced electric current field response for an example drift path near an electrode of interest from the first anode plane in the DUNE horizontal drift detector design.  The left panel shows time domain and the right shows the magnitude of the frequency domain (half) spectrum.  The originally calculated response is in black and red shows the result of LMN-resampling from \qty{100}{ns} to \qty{64}{ns}.}
  \label{fig:frexample}
\end{figure}

The induced current in an anode plane electrode that is due to a nearby electron drift path (e.g. the central row from~\autoref{fig:frs}) is shown in~\autoref{fig:frexample}.
Such a path contributes a large fraction of the total induced current compared to more distant paths.
The $I_s$ shows the original response while $I_r$ shows the LMN-resampling of this sequence.  Note, this resampling is normalized following the interpolation interpretation described in~\autoref{sec:norm}.  

\begin{figure}
  \includegraphics[width=0.9\textwidth,page=4]{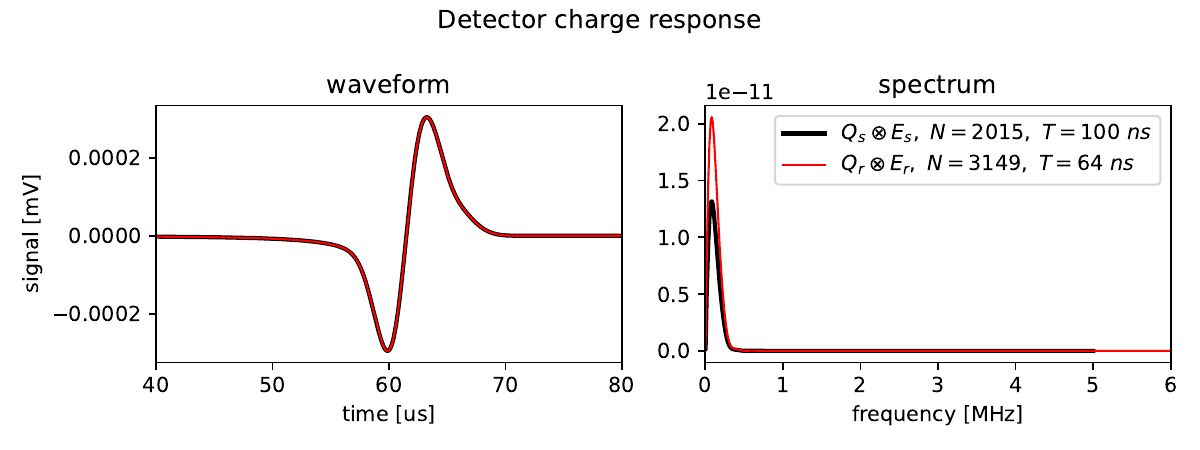}
  \caption{One induced charge field response for an example drift path near an electrode of interest from the first anode plane in the DUNE horizontal drift detector design.  The left panel shows time domain and the right shows the magnitude of the frequency domain (half) spectrum.  The originally calculated induced current field response convolved with the electronics response is shown in black.  The red shows the induced current field response LMN-resampled from \qty{100}{ns} to \qty{64}{ns} convolved with electronics response sampled at \qty{64}{ns}.}
  \label{fig:drexample}
\end{figure}

High frequency ringing can be seen in the resampled field response.  This is due to the fact that the original field response is not fully band limited at \qty{100}{ns}.  This can be seen in the small step down at the original \qty{5}{MHz} Nyquist frequency.  After convolution with the slower electronics response, which is a low-pass filter, in order to form the overall detector response this ringing is naturally abated as illustrated in~\autoref{fig:drexample}, 
The overall detector response is in units of induced electric charge and not induced electric current.
That is, the integrated normalization introduced in~\autoref{sec:norm} is taken by multiplying the current by the sample period. 

Both sets of responses, each at a given sampling, are first input to the simulation in order to output signal-only, voltage-level waveforms. 
While such measurements are not directly observable in a real detector, comparing the native \qty{500}{ns}, \qty{512}{ns}, and the $\qty{512}{ns} \to \qty{500}{ns}$ LMN-resampled waveforms can reveal deviations or artifacts that might otherwise be obscured by realistic noise and ADC integer truncation.
It is important to note that these artifacts are not introduced by the LMN resampling method itself but arise from inherent limitations in the noise and signal simulation.

\begin{figure}
  \includegraphics[width=0.9\textwidth,page=1]{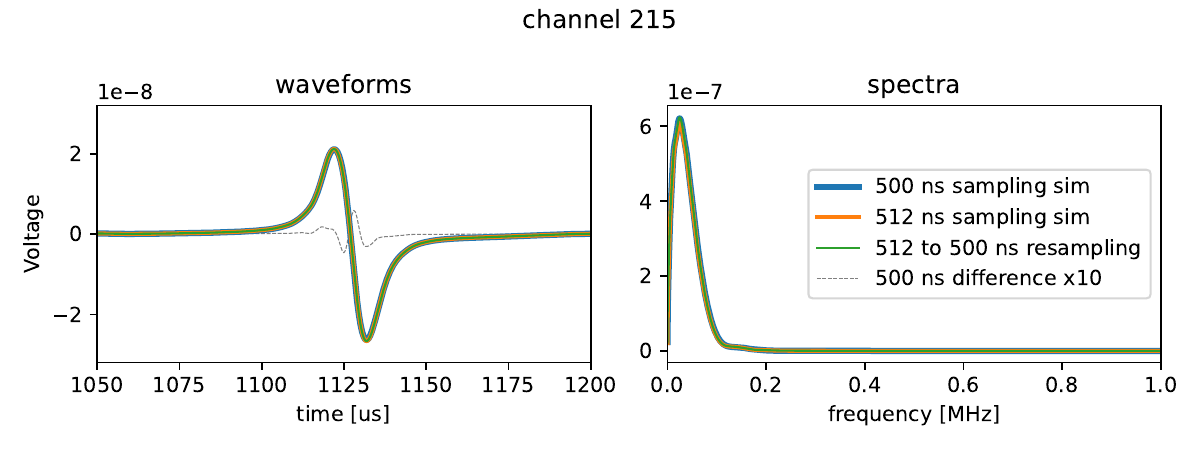}

  \includegraphics[width=0.9\textwidth,page=2]{lartpc/signal-full-pdsp.pdf}

  \includegraphics[width=0.9\textwidth,page=3]{lartpc/signal-full-pdsp.pdf}
  \caption{Examples of a simulated noise-free, signal-only voltage-level waveforms from a select channel of each anode plane.  Two are from otherwise identical simulated events with the first sampled at \qty{500}{ns} (blue) and the second at \qty{512}{ns} (orange).  The third waveform (green) is the result of $\qty{512}{ns} \to \qty{500}{ns}$ LMN resampling.  The gray dashed line represents the difference between the native $\qty{500}{ns}$ sampling and the resampled curves, multiplied by a factor of $10$.}
  \label{fig:signal}
\end{figure}

Examples from one representative channel of each anode plane show very good alignment in the time domain as seen in~\autoref{fig:signal}. 
The noise-free voltage tier, in particular, enables a meaningful comparison between the native and resampled waveforms, as both share a common sampling period, allowing differences to be clearly discernible without being masked by noise.
These differences, magnified by a factor of 10, are shown as gray dashed lines in the left panels of~\autoref{fig:signal}.
A portion of these discrepancies arises from the differential downsampling of the response applied in the simulation, rather than from the LMN resampling method itself.
Additionally, the peaks in the difference curves occur when the original waveforms are rapidly decreasing, making them more likely to be diminished by subsequent signal processing.

\begin{figure}
  \includegraphics[width=0.9\textwidth,page=1]{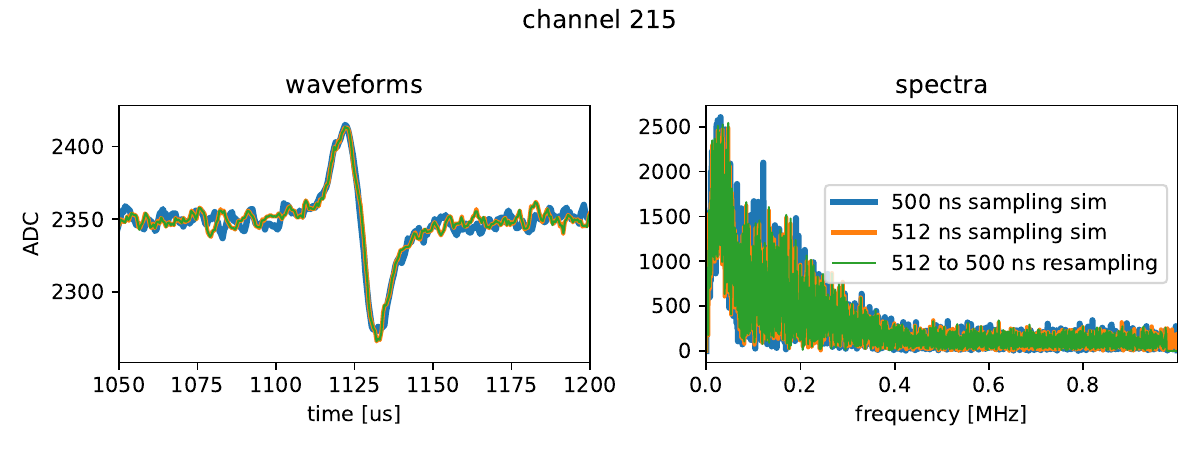}

  \includegraphics[width=0.9\textwidth,page=2]{lartpc/sigdig-full-pdsp.pdf}

  \includegraphics[width=0.9\textwidth,page=3]{lartpc/sigdig-full-pdsp.pdf}
  \caption{The voltage-level signal waveforms of~\autoref{fig:signal} after noise is added and the ADC digitizer model is applied.  The \qty{500}{ns} (blue) and \qty{512}{ns} (orange) sampling cases are shown.  The third waveform (green) is the result of resampling the \qty{512}{ns} ADC-level waveform to \qty{500}{ns}.}
  \label{fig:sigdig}
\end{figure}

Waveforms measured by real detectors include signal and noise.
In the simulation these two components are added at the voltage level and then scaled and truncated to integer by the ADC.
These waveforms for native \qty{500}{ns} and \qty{512}{ns} simulation as well as the $\qty{512}{ns} \to \qty{500}{ns}$ LMN-resampling at ADC level are shown in ~\autoref{fig:sigdig}. 
The detailed noise components in the \qty{512}{ns} waveform are closely followed in the $\qty{512}{ns} \to \qty{500}{ns}$ resampling.
It can also be seen that the main signal body is faithfully resampled, well within the variation due to differently generated noise between the native \qty{500}{ns} simulation and the noise shared by the native \qty{512}{ns} simulation and the $\qty{512}{ns} \to \qty{500}{ns}$ resampling.

\begin{figure}
  \includegraphics[width=0.9\textwidth,page=1]{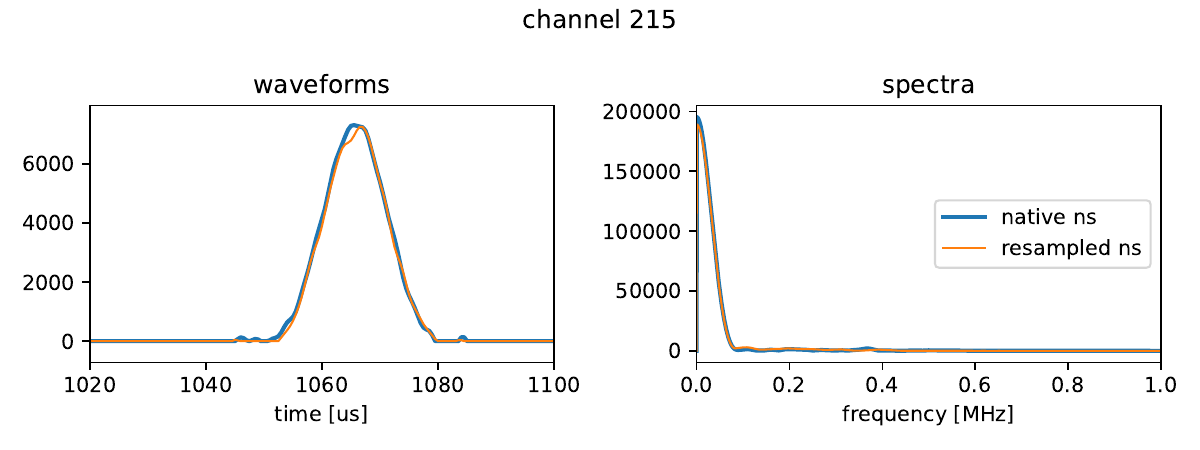}

  \includegraphics[width=0.9\textwidth,page=2]{lartpc/sigproc-full-pdsp.pdf}

  \includegraphics[width=0.9\textwidth,page=3]{lartpc/sigproc-full-pdsp.pdf}
  \caption{The output of the signal processing applied to native \qty{500}{ns} ADC-level simulation and the \qty{512}{ns} ADC-level simulation after resampling to \qty{500}{ns}.}
  \label{fig:sigproc}
\end{figure}

The 2D signal processing described in~\autoref{sec:sp} is sensitive to spectral distortions.  And so, as a final comparison, signal processing using the responses corresponding to the \qty{500}{ns} sampling is applied to the native \qty{500}{ns} ADC-level signal and noise simulation and the $\qty{512}{ns} \to \qty{500}{ns}$ resampling.  The results from the same channels selected above are shown in~\autoref{fig:sigproc}.  The small variations are due to the noise components that differ between the two waveforms.

With confidence that the resampling preserves information about the underlying
signal implementation issues can now be considered. 
To estimate the absolute performance, LMN resampling was applied to readouts of
6000 samples across 2560 channels which is a typical size for DUNE.  
About 300 CPU core-millisecond were required to resample this size readout. 
As a comparison, $\mathcal{O}(\qty{10})$ CPU core-seconds are required to apply
the subsequent signal processing stage to the readout.
Future work may accelerate the signal processing via GPU and may achieve
$10\times$ or better speedup.
The resampling is well suited for GPU acceleration and it may be accelerated in
order to avoid becoming a bottleneck to the accelerated signal processing.
Using the relative performance measures given in \autoref{sec:cinoare} we may
estimate the naive method may require aout $10\times$ more CPU time than LMN. 
While that is still competitive compared to the CPU time required for signal
processing, the inflation to achieve the ``naive'' integer ratio would require
$\mathcal{O}(100\times)$ more memory.

\section{Summary}\label{sec:Summ}
In this paper we describe the LMN method which achieves a fast and robust waveform resampling based on the Fast Fourier Transformation algorithm.
The principle of the LMN method is introduced.
It exploits freedom to resize the original signal to enable a special case optimization that in turn allows for upsampling and downsampling by integer factors to be performed simultaneously in the frequency domain.
Intermediate signals are kept to modest sizes and thus ruinous computational costs are avoided.
Normalization for a number of different signal interpretations is provided.
The potential for a resampling to produce artifacts is raised and mitigation procedures enabled by the method are outlined.
The performance in terms of quality and computational cost of the LMN method is compared with some popular methods.
The LMN method is generally among the fastest and most accurate.
By exercising the simulation and signal processing procedures used for Liquid Argon Time Projection Chambers the effectiveness of the LMN method is demonstrated in a particle physics experiment. 

\acknowledgments
We would like to thank Laura Zambelli for her insightful comments and valuable suggestions during the DUNE collaboration review of this paper.

\bibliography{mybib}

\end{document}